\documentclass[aps,prl,amsmath,amssymb,floatfix,twocolumn,amsmath,superscriptaddress,twocolumn,nofootinbib,tighten,letterpaper]{revtex4}
\usepackage{multirow}
\usepackage{bbold}
\usepackage{subfigure}
\usepackage{color}
\usepackage{mathrsfs}
\usepackage{hyperref}
\usepackage[normalem]{ulem}
\usepackage{bm}

\usepackage{amsfonts, relsize, color}
\usepackage{graphics}
\usepackage{graphicx}
\usepackage{subfigure}
\usepackage{hyperref}
\usepackage{color}
\usepackage{comment}

\begin{document}
\title{Inner Skin Effects on Non-Hermitian Topological Fractals}

\author{Sourav Manna}
\affiliation{Department of Condensed Matter Physics, Weizmann Institute of Science, Rehovot 7610001, Israel}

\author{Bitan Roy}\thanks{Corresponding author:bitan.roy@lehigh.edu}
\affiliation{Department of Physics, Lehigh University, Bethlehem, Pennsylvania, 18015, USA}

\date{\today}

\maketitle

\noindent
{\bf Abstract}\\
{\bf Non-Hermitian (NH) crystals, quasicrystals and amorphous network display an accumulation of a macroscopic number of states near one of its specific interfaces with vacuum, such as edge, surface, hinge or corner. This phenomenon is known as the NH skin effect, which can only be observed with open boundary condition. In this regard self-similar fractals, manifesting inner boundaries in the interior of the system, harbor a novel phenomenon, the \emph{inner skin effect} (ISE). Then the NH skin effect appears at the inner boundaries of the fractal lattice with periodic boundary condition. We showcase this observation by implementing prominent models for NH insulators and superconductors on representative planar Sierpinski carpet fractal lattices. They accommodate both first-order and second-order ISEs at inner edges and corners, respectively, for charged as well as neutral Majorana fermions. Furthermore, over extended parameter regimes ISEs are tied with nontrivial bulk topological invariants, yielding intrinsic ISEs. With the recent success in engineering NH topological phases on highly tunable metamaterial platforms, such as photonic and phononic lattices, as well as topolectric circuits, the proposed ISEs can be observed experimentally at least on fractal metamaterials with periodic boundary condition.}      

\noindent 
{\bf Introduction}\\
Symmetries play a pivotal role in structural classification of solids. For example, crystals typically exhibit three discrete symmetries: rotation, translation and reflection, which are, however, absent in amorphous materials. On the other hand, quasicrystals manifest crystal-forbidden discrete rotational symmetries, such as the eight-fold one on Ammann-Beenker tiling~\cite{janot:QCbook}. Yet another class of systems, fractals possess a unique symmetry, namely self-similarity, leading to pattern repetition over many scales~\cite{fractal:book}. Often natural objects, such as biological cells, coastlines, trees etc. display (approximate) self-similarity. As its direct consequence fractals feature inner boundaries in the interior of the systems. With recent realizations of quantum fractals in designer electronic~\cite{cmsmith2019:frac} and molecular~\cite{wu2015:frac} materials, unique signatures of inner boundaries on quantum phenomena have become a timely issue of pressing fundamental importance~\cite{neupert2018:frac, spaiprem2019:frac, katsnelson2020:frac, souravmanna2020:frac, larsfirtz2020:frac, segev2020:frac, souravmanna2021:frac, mannanandyroy2021:frac, mannajaworowskinielsen2021:fractal, ivakietal2021:fractal}, bearing direct experimental pertinence.

Throughout the topological age of condensed matter physics, boundaries (such as edge, surface, hinge and corner) have served as litmus probes for the experimental detection of topological materials. Namely, gapless modes appear at these interfaces, when electronic wavefunctions feature nontrivial geometry in the bulk of any material: the bulk-boundary correspondence~\cite{Hasan-Kane-RMP, Qi-Zhang-RMP}. This concept also extends to the landscape of classical metamaterials, such as topological photonic~\cite{ozawa-rmp2019} and phononic~\cite{susstrunk-science2015, yang-prl2015} lattices, as well as topolectric circuits~\cite{ninguyan-prx2015, albert-prl2015, imhof-natphys2018}.

\begin{figure}[t!]
\includegraphics[width=0.92\linewidth]{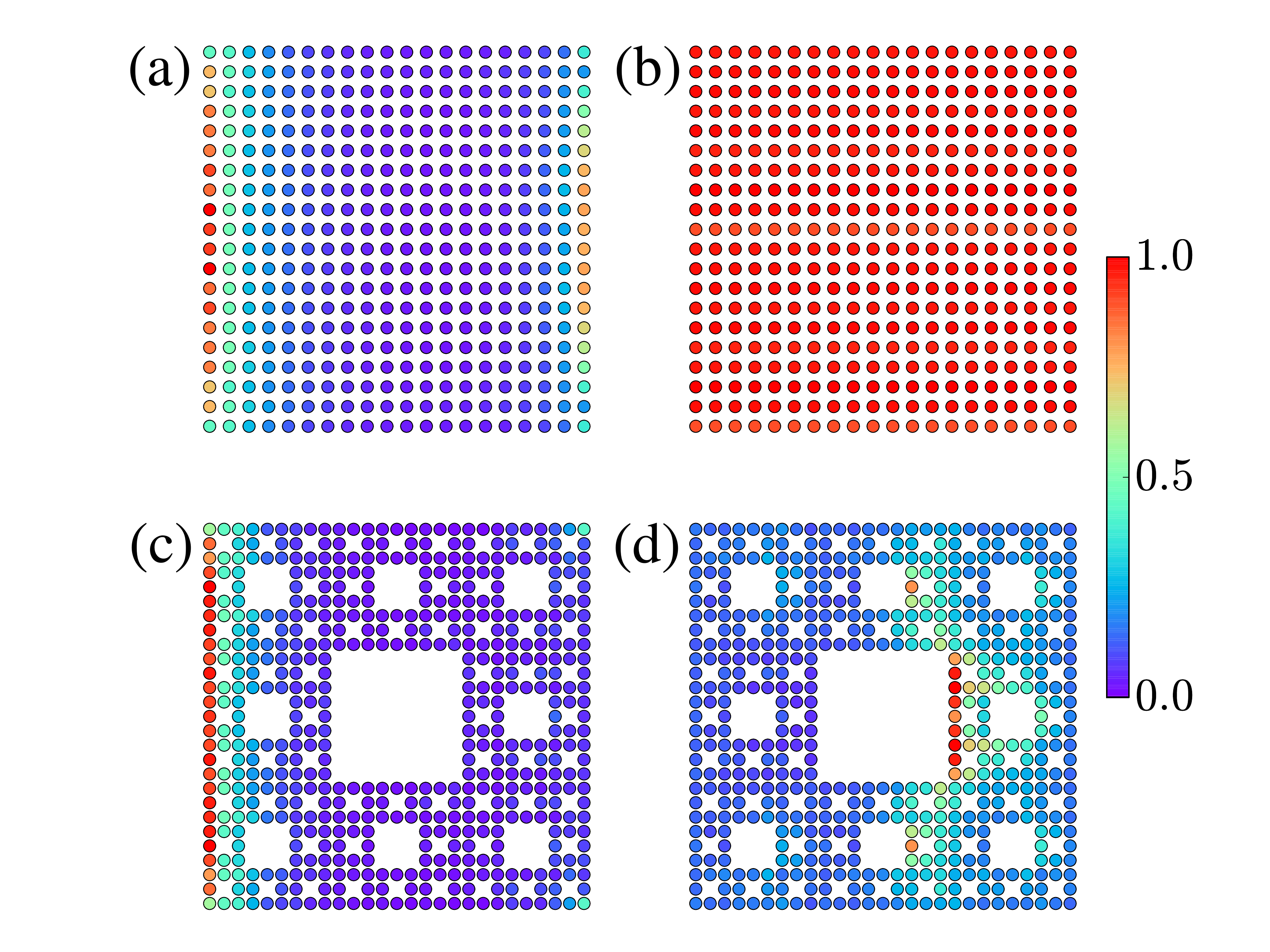}
\caption{{\bf First-order non-Hermitian (NH) skin effect on a square lattice and Sierpinski carpet (SC) fractal}. (a) Total weight of all the left eigenvectors (normalized by its maximum value) of $H_{\rm FO}$ [Eq.~(\ref{eq:HamilNHfirstorder})] on a square lattice with open boundary condition (OBC), displaying NH skin effect at outer left edge. (b) Same as (a) but with periodic boundary condition (PBC) in the $x$ direction, showing no skin effect. (c) Same as (a), but on a SC fractal of third generation (containing 512 sites) with OBC, showing NH skin effect at its outer left edge. (d) Same as (c), but with PBC only in the $x$ direction, confirming the proposed NH inner skin effect at its inner left edges. The outcomes in (b) and (d) remain qualitatively unchanged with additional PBC in the $y$ direction. Here we set $m_0=0$, ${\boldsymbol h}=(0.4,0,0)$, $t=t_0=1$, $r_0=a$ and $R=2 a$, where $a$ is the distance between the nearest-neighbor sites [Eq.~(\ref{eq:hoppingconfiguration})]. For right eigenvectors the skin effect appears on the right edges. See Supplementary Fig.~1.      
}~\label{fig:Fig1}
\end{figure}

\begin{figure}[t!]
\includegraphics[width=0.90\linewidth]{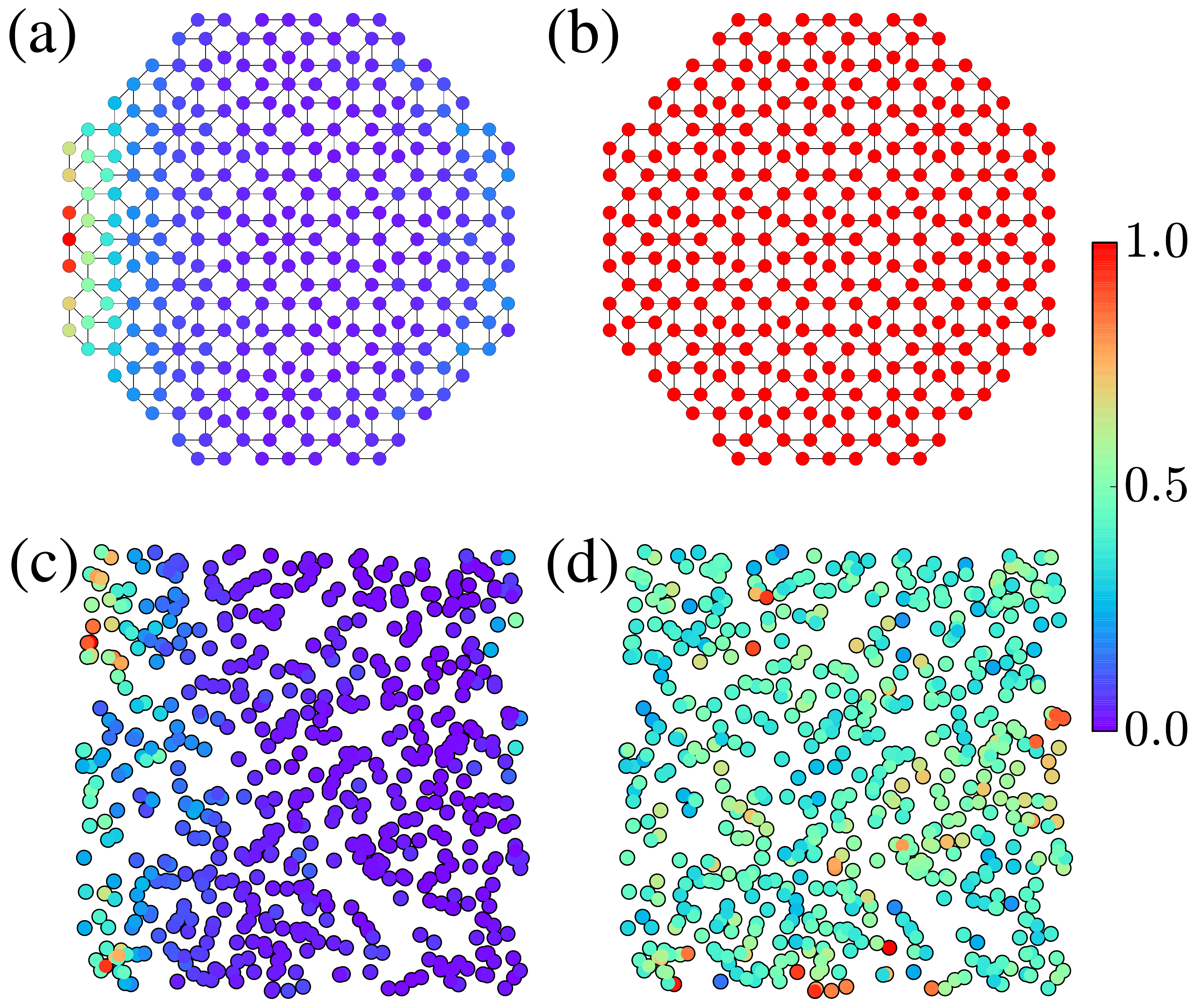}
\caption{{\bf Skin effect on Ammann-Beenker quasicrystal and amorphous network.} (a) Non-Hermitian skin effect for the left eigenvectors of $H_{\rm FO}$ on the left edge of an octagonal Ammann-Beenker quasicrystal with open boundary condition (OBC) for $t=t_0=1$, $m=1.0$, $h_x=0.75$, $R=2 a$ and $r_0=a$, where $a$ is the length of its each arm (black line). (b) Same as (a), but with periodic boundary condition (PBC) in the $x$ direction, showing no skin effect. (c) Same as (a), but on an amorphous network of 600 lattice sites for $h_x=0.4$ and $R=0.98 a$. Here $a$ is the linear dimension of the network in each direction. Other parameters are kept unchanged. (d) Same as (c), but with PBC along the $x$ direction. A few isolated and randomly distributed points on the network show a slightly higher densities. But, after averaging over many (typically 250) realizations, we find that there is no skin effect in the bulk of an amorphous network with PBC. Right eigenvectors show skin effects on the right edges of both systems with OBC. See Supplementary Figs.~1-3.         
}~\label{fig:Fig2}
\end{figure}

When topological materials interact with the environment, which can be modeled by non-Hermitian (NH) operators, the weight of a macroscopic number of states shifts toward a specific boundary, a phenomenon known as the NH skin effect. On crystals NH skin effect can be observed only with open boundary condition (OBC) [Fig.~\ref{fig:Fig1}(a)], while it is absent with periodic boundary condition (PBC) [Fig.~\ref{fig:Fig1}(b)]~\cite{torres:review, tanmoy:review, Bergholtz:review, Kohmoto:PRB2011, Huang:PRA2013, LFu:arXiv2017, torres:PRB2018, Wang:PRL2018A, Wang:PRL2018B, ueda:PRX2018, ueda:PRB2018, Liangfu2018, Bergholtz:PRL2018, Ueda:NatComm2019, Murakami:PRL2019, LeeThomale:PRB2019, Sato:PRL2020, Vitelli:PRL2020, Xue:NatPhys2020, Fang:PRL2020, Slager:PRL2020, Song:PRB2020, hughes:PRL2021, panigrahi:2021, schindlerprem:2021, Moghaddam:PRB2021, roccati:PRA2021}. Here we unfold novel \emph{inner skin effect} (ISE), exclusively available on fractal networks with PBCs, manifesting prominent skin effects at its inner boundaries or skins, by implementing key NH models on the Sierpinski carpet (SC) fractal lattices [Figs.~\ref{fig:Fig1}(c) and \ref{fig:Fig1}(d)], characterized by the fractal dimension $d_{\rm frac}=1.89$. By contrast, the fate of NH skin effect on quasicrystals and amorphous lattices are qualitatively similar to that on the square lattice [Fig.~\ref{fig:Fig2}].

\begin{figure}[t!]
\includegraphics[width=0.90\linewidth]{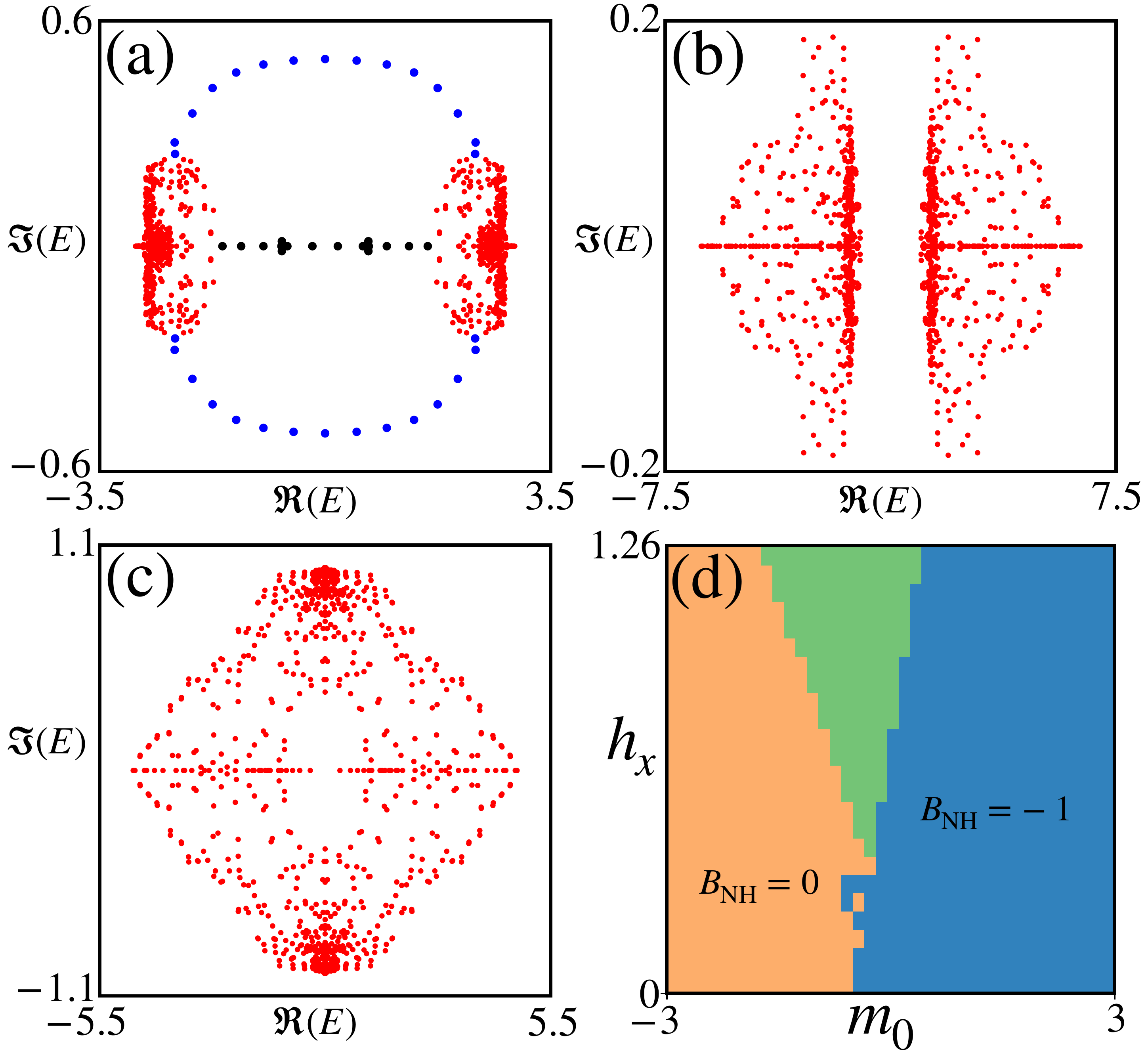}
\caption{{\bf Energy spectra and phase diagram on non-Hermitian (NH) Sierpinski carpet fractal}. The energy spectra of $H_{\rm FO}$ [Eq.~(\ref{eq:HamilNHfirstorder})] for $t=t_0=1$, $r_0=a$, $R=2 a$, and (a) $m_0=2.0$ and $h_x=0.5$, (b) $m_0=-2.0$ and $h_x=0.4$ and (c) $m_0=-0.5$ and $h_x=1.0$ with periodic boundary condition (PBC) only in the $x$ direction. The system describes a NH (a) Chern insulator with the NH Bott index $B_{\rm NH}=-1$, (b) trivial insulator with $B_{\rm NH}=0$, and (c) insulator, where $B_{\rm NH}$ is not well defined. In (a) and (b) the spectra display line gaps. In (c) the spectra display a point gap. In (a) two spectral lobes with $\Re(E)<0$ and $\Re(E)>0$ are connected via topological modes, localized at top and bottom outer edges (blue dots) and inner edges (black dots). (d) Phase diagram of $H_{\rm FO}$ in the $(m_0,h_x)$ plane hosts these three insulating phases with $B_{\rm NH}=-1$ (blue region), $B_{\rm NH}=0$ (orange region) and where $B_{\rm NH}$ is not well defined (green region). These outcomes remain qualitatively unchanged when we impose additional PBC in the $y$ direction, except the outer edge modes [blue dots in (a)] disappear. See Supplementary Fig.~4.        
}~\label{fig:Fig3}
\end{figure}

First, we consider the NH Chern model, which when features skin effect at the outer left edge of SC fractal lattice with OBCs, the ISE appears at inner left edges upon implementing PBC only along the $x$ direction. The ISE in this case is first-order in nature, as a one-dimensional edge on a two-dimensional SC fractal lattice is characterized by the codimenion $d_c=2-1=1$. Although first-order ISE is observed for an arbitrary strength of the NH coupling, for its small values the system features line gaps about zero energy and two distinct insulators with NH Bott indices $B_{\rm NH}=-1$ and $0$. The former phase thus exhibits intrinsic first-order ISE as it is tied with a nontrivial bulk topological invariant [Fig.~\ref{fig:Fig3}].

\begin{figure}[t!]
\includegraphics[width=0.95\linewidth]{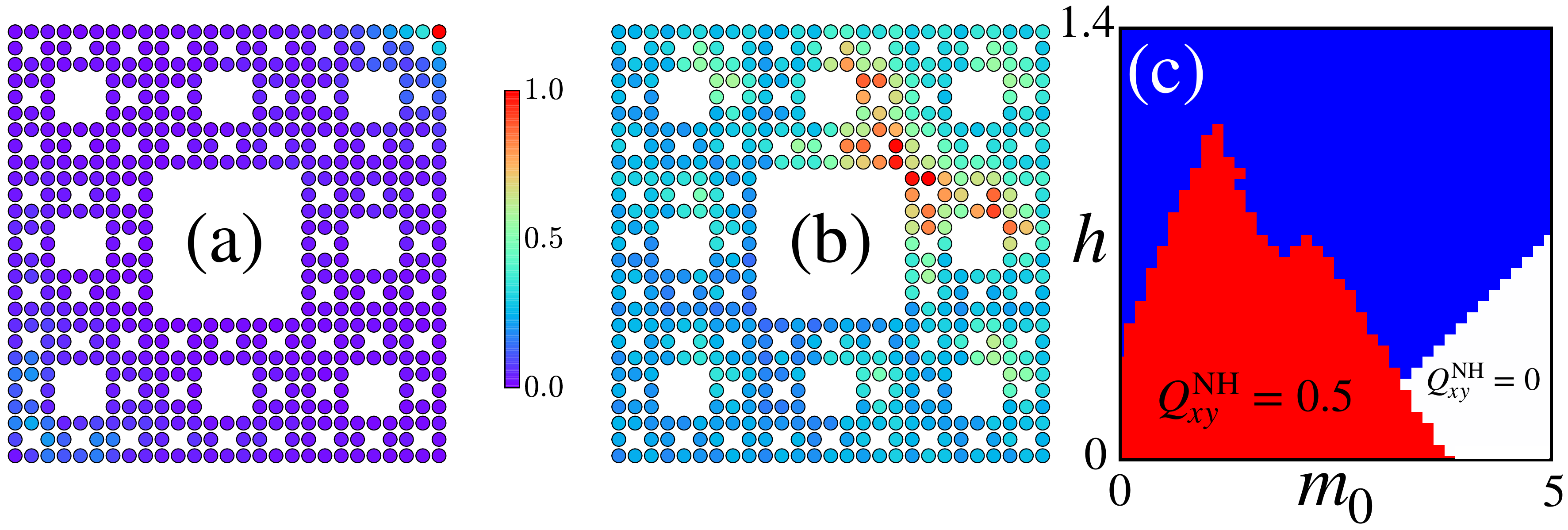}
\caption{{\bf Non-Hermitian (NH) second-order skin effect and associated phase diagram on Sierpinski carpet fractal.} (a) Total weight of all the left eigenvectors (normalized by its maximum value) of $H_{\rm SO}$ [Eq.~(\ref{eq:HamilNHsecondorder})] for $m_0=1.5$, $g=0.5$ and ${\boldsymbol h}=(0.35,0.35,0)$ with open boundary conditions in all directions, showing second-order skin effect at an outer corner. (b) Same as (a), however with periodic boundary condition in both $x$ and $y$ directions, displaying a second-order inner skin effect around inner corners. (c) Global phase diagram of $H_{\rm SO}$ on the $(m_0, h)$ plane with the same boundary condition as in (b) and $g=0.5$, featuring (i) a NH quadrupole insulator with NH quadrupole moment $Q^{\rm NH}_{xy}=0.5$ (red region), (ii) a NH trivial insulator with $Q^{\rm NH}_{xy}=0.0$ (white region), and (iii) a NH insulator where $Q^{\rm NH}_{xy}$ is not well defined (blue region). Throughout we set $t=t_0=1$, $R=2 a$ and $r_0=a$. The right eigenvectors of $H_{\rm SO}$ display second-order (inner) skin effect around the diagonally opposite corners, as shown in Supplementary Fig.~5.       
}~\label{fig:Fig4}
\end{figure}

\begin{figure}[t!]
\includegraphics[width=0.95\linewidth]{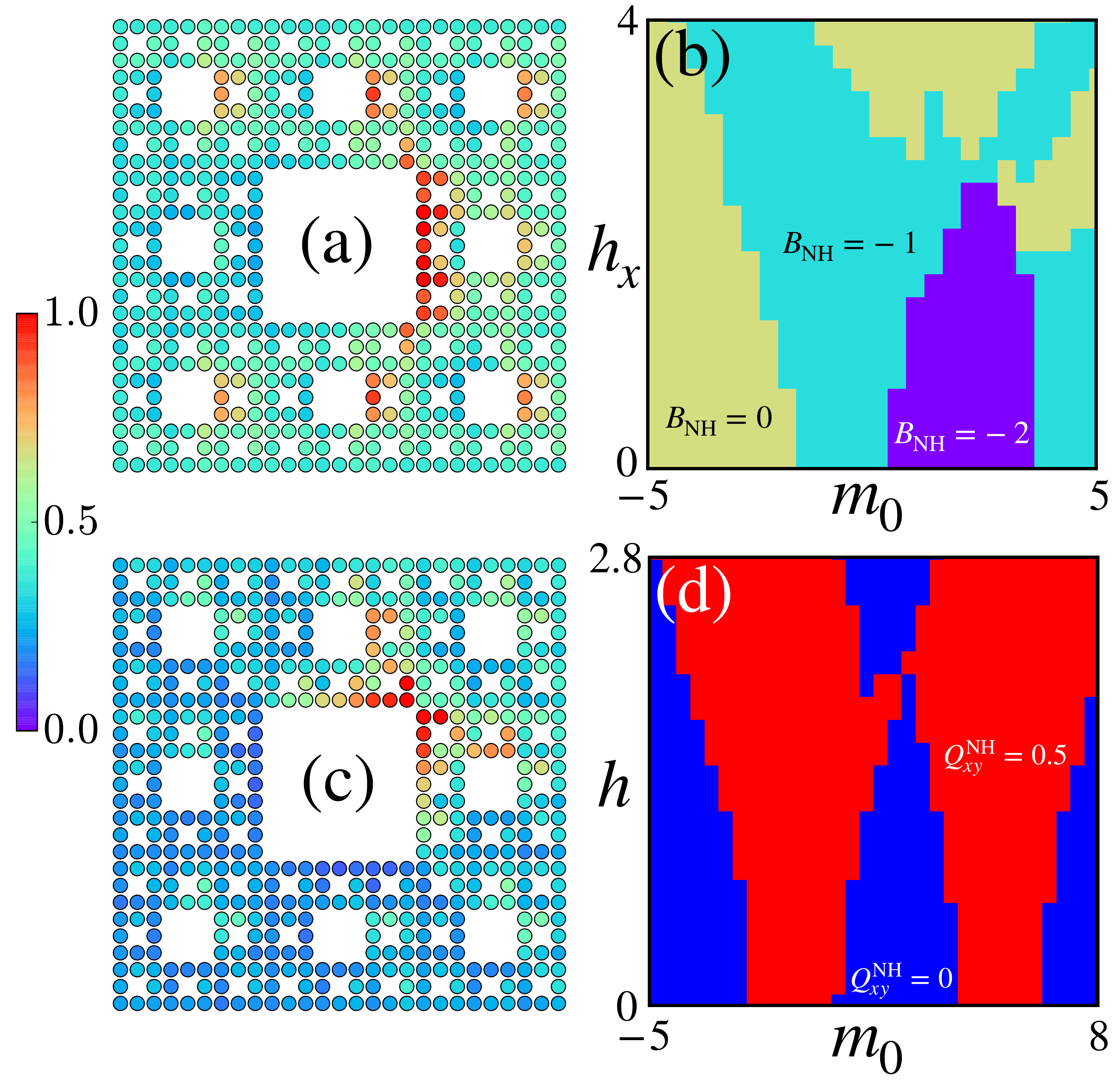}
\caption{{\bf Mojorana inner skin effects (ISEs) and associated global phase diagrams}. (a) First-order Majorana ISE for the left eigenvectors of $H^{\rm pair}_{\rm FO}$ [Eq.~(\ref{eq:Majoranafirstorder})] when $m_0=0.0$, $\Delta_{\rm FO}=0.25$ and ${\boldsymbol h}=(0.5,0,0)$. (b) Global phase diagram of $H^{\rm pair}_{\rm FO}$ on the $(m_0,h_x)$ plane for $\Delta_{\rm FO}=1.0$, constructed from the non-Hermitian (NH) Bott index $B_{\rm NH}$. (c) Second-order Majorana ISE for the left eigenvectors of $H^{\rm pair}_{\rm SO}$ [Eq.~(\ref{eq:Majoranasecondorder})] when $m_0=3.0$, $g=0.5$, $\Delta_{\rm SO}=0.025$ and ${\boldsymbol h}=(-0.53,-0.53,0)$. (d) Global phase diagram of $H^{\rm pair}_{\rm SO}$ on the $(m_0,h)$ plane, constructed by computing the NH quadrupole moment $Q^{\rm NH}_{xy}$ for neutral Majorana fermions for $g=0.5$ and $\Delta_{\rm SO}=1.25$. Throughout we set $t=t_0=1$, $R=8 a$ and $r_0=a$. Right eigenvectors of $H^{\rm pair}_{\rm FO}$ ($H^{\rm pair}_{\rm SO}$) show inner first-order (second-order) skin effects, but around opposite edges (diagonally opposite corners), as shown in Supplementary Fig.~6.          
}~\label{fig:Fig5}
\end{figure}

SC fractal also accommodates second-order ISE, where the weight of a macroscopic number of states is highly localized predominantly around an inner corner with $d_c=2-0=2$ [Fig.~\ref{fig:Fig4}]. Namely, when the skin effect is observed at one of the four outer corners of the SC fractal with OBCs, it shifts toward inner corners when the PBC is introduced along both $x$ and $y$ directions. For small NH couplings, the system features quantized NH quadrupole moment $Q^{\rm NH}_{xy}=0.5$, displaying intrinsic second-order ISE, besides the trivial one $Q^{\rm NH}_{xy}=0.0$. For large NH couplings when $Q^{\rm NH}_{xy}$ is no longer well defined, the system still displays second-order ISEs. Finally, we show that when NH SC fractal hosts topological pairings, both first- and second-order ISEs for neutral Majorana fermions are observed with appropriate PBCs [Fig.~\ref{fig:Fig5}].

\noindent 
{\bf Results}\\
{\bf First-order ISE}.~First-order ISE on the SC is captured by the model Hamiltonian $H_{\rm FO}=H_0 + H_{\rm NH}$, where 
\begin{eqnarray}~\label{eq:HamilNHfirstorder}
\hspace{-0.5cm}H_0 &=&\sum_{j \neq k} \frac{F(r_{jk})}{2} c^\dagger_j \big[  -it (  \cos\phi_{jk}\Gamma_1 + \sin\phi_{jk} \Gamma_2) \nonumber \\
&+& t_0 \Gamma_3  \big] c_k - \sum_{j} c_j^\dagger \big[m_0 \Gamma_3 \big] c_j, \nonumber \\
\hspace{-0.5cm} H_{\rm NH} &=& i \; c_j^\dagger ({\boldsymbol h} \cdot {\boldsymbol \Gamma}) c_j \equiv i \; c_j^\dagger \big[ h_x \Gamma_1 + h_y \Gamma_2 + h_z \Gamma_3 \big] c_j, 
\end{eqnarray} 
$i=\sqrt{-1}$, $c_j=[c_{j\alpha},c_{j\beta}]^\top$, $c_{j\alpha}$ is the fermionic annihilation operator at site $j$ and on orbital $\alpha$, and $\Gamma_\mu=\tau_\mu$ for $\mu=1,2,3$. The vector Pauli matrix ${\boldsymbol \tau}$ operates on the orbital index. Hopping strength between sites $j$ and $k$, respectively placed at ${\bf r}_j$ and ${\bf r}_k$, is augmented by  
\begin{equation}~\label{eq:hoppingconfiguration}
	F(r_{jk}) = \Theta(r_{jk}-R) \exp \left[1 - \frac{r_{jk}}{r_0} \right],
\end{equation}
ensuring that the sites are well connected, especially in the absence of translational symmetry. Here $r_{jk} = |{\bf r}_j - {\bf r}_k|$ is the distance and $\phi_{jk}$ is the azimuthal angle between them, $R$ controls the range of hopping, and $r_0$ is the decay length. Generic momentum-independent NH coupling is captured by $H_{\rm NH}$. This model can be implemented on a square lattice, quasicrystals, amorphous netwrok~\cite{agarwalashenoy} as well as on a SC fractal lattice~\cite{mannanandyroy2021:frac}. In the Supplementary Note 1 we show that $H_0$ can be derived from the spinless Bernevig-Hughes-Zhang model~\cite{bhz-science2006}. Consult also Supplementary Table 1.

On a square lattice with OBC, the above model shows NH skin effect on the outer left edge for ${\boldsymbol h} =(h_x,0,0)$ [Fig.~\ref{fig:Fig1}(a)]. With PBC in the $x$ direction or both $x$ and $y$ directions, there is no skin effect anywhere in the system [Fig.~\ref{fig:Fig1}(b)].  When ${\boldsymbol h}=(0,h_y,0)$, the role of the $x$ and $y$ directions get reversed. By contrast, for ${\boldsymbol h}=(0,0,h_z)$ there is no skin effect, irrespective of the boundary condition. So, we set $h_z=0$ for the rest of the discussion. On the SC with OBC, the situation is similar to that on a square lattice [Fig.~\ref{fig:Fig1}(c)]. However and most importantly, the situation changes dramatically when we impose PBC only in the $x$ direction with ${\boldsymbol h}=(h_x,0,0)$. The weight of a macroscopic number of states then piles up at the inner left edges, extended in the $y$ direction of the fractal lattice, manifesting the first-order ISE [Fig.~\ref{fig:Fig1}(d)]. With ${\boldsymbol h}=(0,h_y,0)$, the role of the inner edges extended along the $x$ and $y$ directions and the requisite periodic boundary conditions in these two directions are reversed. The NH skin effect on Ammann-Beenker quasicrystal and amorphous network with open and periodic geometries are qualitatively similar to those on the square lattice [Fig.~\ref{fig:Fig2}]. The NH skin effect on crystals in the presence of random on site impurities are similar to that in amorphous materials, plagued by structural disorder~\cite{hughes:PRB2021disorder}. Also a NH impurity coupling of varying strength between two sites residing at the end of a one-dimensional chain can display conventional NH skin effect, regular and reverse scale free accumulations~\cite{leegong:commphys2021}. However, with PBC such system does not display any NH skin effect.

To relate the first-order ISE to a bulk topological invariant, we compute the NH Bott index ($B_{\rm NH}$). To this end we scale the coordinates of all the sites $(x_j,y_j)$ belonging to the SC fractal by its linear dimension in these two directions, such that $x_j \in [0,1]$ and $y_j \in [0,1]$ for all $j$. We define two unitary matrices $U_x = \exp(2\pi i X )$ and $U_y = \exp(2\pi i Y)$, where the elements of the diagonal matrices $X$ and $Y$ are $X_{j,k}=x_j \delta_{jk}$ and $Y_{j,k}=y_j \delta_{jk}$, respectively. Here $\delta_{jk}$ is the Kronecker delta. Four choices of the projector to the half-filled system are 
\begin{equation}
	P \in \sum_{\Re(E) < 0} \Big\{ | R_E \rangle \langle L_E |,  | L_E \rangle \langle R_E |, | R_E \rangle \langle R_E |, | L_E \rangle \langle L_E | \Big\}, \nonumber
\end{equation}
where $| R_E \rangle$ and $| L_E \rangle$ are the right and left eigenstates with energy $E$ (typically complex), respectively. Then 
\begin{equation}~\label{eq:NHbottindex}
		B_{\rm NH} = \frac{1}{2 \pi} \text{Im} \left[ \text{Tr} \left\{ \ln \Big[    V_x V_y V_x^\dagger V_y^\dagger  \Big] \right\} \right],  
\end{equation}
where $V_\ell = I - P + P U_\ell P$ for $\ell=x$ and $y$~\cite{bottindex1, bottindex2}.

For small $h_x$ (or $h_y$), the system supports two distinct NH insulators with $B_{\rm NH}=-1$ (topological) and $0$ (trivial). The energy spectra of $H_{\rm FO}$ in these two phases display line gaps [Fig.~\ref{fig:Fig3}(a) and \ref{fig:Fig3}(b)], which can be smoothly connected to the normal band gap in a corresponding Hermitian system. Especially, when $B_{\rm NH}=-1$, the associated ISE is intrinsic in nature, tied with a nontrivial topological invariant. For large $h_x$, $B_{\rm NH}$ ceases to be a bonafide topological invariant. In this regime, the energy spectra show point gap [Fig.~\ref{fig:Fig3}(c)], which cannot be smoothly connected to a Hermitian band gap. In crystals, such parameter regime can support gapless exceptional points~\cite{torres:review, tanmoy:review, Bergholtz:review}, which, however, we could not find on the SC fractal, possibly due to the lack of translational symmetry. Finally, in terms of $B_{\rm NH}$ we construct a global phase diagram of the model Hamiltonian $H_{\rm FO}$, which is identical for all four choices of $P$ [Fig.~\ref{fig:Fig3}(d)].

Throughout this work we define the line and point gaps with respect to zero energy in half-filled systems, as all the models possess particle-hole symmetry. For example, the particle-hole symmetry of $H_{\rm FO}$ [Eq.~(\ref{eq:HamilNHfirstorder})] is generated by $\Theta_{\rm FO}=\tau_1 {\mathrm I}_{\mathcal A} {\mathcal K}$, as $\{\Theta_{\rm FO}, H_{\rm FO} \}=0$ when $h_z=0$, where ${\mathrm I}_{\mathcal A}$ is a ${\mathcal A}$-dimensional identity matrix, ${\mathcal A}$ is the number of lattice sites in the system and ${\mathcal K}$ is complex conjugation. As a result whenever the energy spectra changes from the line to point gap, the topological invariant, such as $B_{\rm NH}$, changes. In principle, one can define line and point gaps about arbitrary non-zero energy~\cite{satoetal:PRX2019}. But, then line to point gap transition is not necessarily associated with any change of topological invariant. For NH insulators one can introduce a particle-hole symmetry breaking term, which causes mere overall shift of all energy eigenvalues. Then line and point gaps need to be defined about the shifted zero energy, and our prescription holds. By contrast, for NH pairing models the particle-hole symmetry is exact.

{\bf Second-order ISE}.~Having established the first-order ISE, we now solely focus on the SC fractal, which when supports conventional second-order NH skin effect at one of its four outer corners with OBC [Fig.~\ref{fig:Fig4}(a)], a second-order ISE appears with PBCs in both $x$ and $y$ directions for which the weight of a macroscopic number of states accumulate around its inner corners [Fig.~\ref{fig:Fig4}(b)]. To this end, we implement the following NH model Hamiltonian on the SC fractal
\begin{equation}~\label{eq:HamilNHsecondorder}
H_{\rm SO}=H_{\rm FO} + g \; \sum_{j \neq k} \frac{F(r_{jk})}{2} c^\dagger_j \Big[ \; \cos(2\phi_{jk}) \; \Gamma_4  \Big] c_k,   
\end{equation}    
where $c_j = [c_{j \uparrow \alpha}, c_{j \uparrow \beta}, c_{j \downarrow \alpha}, c_{j \downarrow \beta} ]^\top$, and $c_{j \sigma \alpha}$ is the fermionic annihilation operator at site $j$, with spin projection $\sigma=\uparrow, \downarrow$ and on orbital $\alpha$. Hermitian $\Gamma$ matrices are now four-dimensional, given by $\Gamma_1=\sigma_3 \tau_1$, $\Gamma_2=\sigma_0 \tau_2$, $\Gamma_3=\sigma_0 \tau_3$ and $\Gamma_4=\sigma_1 \tau_1$. The newly introduced Pauli matrices $\{ \sigma_\mu \}$ with $\mu=0, \cdots, 3$ operate on the spin indices. The term proportional to $g$ breaks the four-fold rotational symmetry, under which $\phi_{jk} \to \phi_{jk}+\pi/2$. It is responsible for the second-order topology, with the hallmark corner localized zero-energy modes, when ${\boldsymbol h}=0$. One can derive $H_{\rm SO}$ from the Benalcazar-Bernevig-Hughes model~\cite{BBH, royantiunitary}, in the presence of NH couplings, as shown in the Supplementary Note 2. Consult also Supplementary Table 1.

The above model displays second-order skin effect at an outer corner of the SC fractal for ${\boldsymbol h}=(\pm h, \pm h,0)$ with $h>0$, when we impose OBC in all directions [Fig.~\ref{fig:Fig4}(a)], similar to the situation on a square lattice. However, when PBCs are imposed in both $x$ and $y$ directions, the second-order skin effect shifts toward the inner corners of the fractal, yielding a second-order ISE [Fig.~\ref{fig:Fig4}(b)]. A couple of comments are due at this stage. As any corner is equally shared by the edges, extended in the $x$ and $y$ directions, PBC must be imposed in these two directions in order to observe the second-order ISE. Furthermore, the coordination number of an outer corner is two. But, such a corner cannot be found in the interior of the SC fractal. Consequently, second-order ISE spreads slightly away from inner corners. For this model the particle-hole symmetry is generated by $\Theta_{\rm SO}=\sigma_3 \tau_1 {\mathrm I}_{\mathcal A} {\mathcal K}$.

The second-order ISE can be related to the NH quadrupole moment $Q^{\rm NH}_{xy}=n-n_{\rm al}$ (modulo 1), where
\begin{eqnarray}
\hspace{-0.5cm} n=\Re \left[ \frac{i}{2 \pi} {\rm Tr} \left( \ln \left\{ U^\dagger_j  \exp \left[ 2 \pi i \sum_{\bf r} \hat{q}_{xy} ({\bf r}) \right]  U_k \right\} \right) \right],
\end{eqnarray}
and $\hat{q}_{xy} ({\bf r})= x y/\ell^2$~\cite{agarwala:octupolar, hughes:octupolar, cho:octupolar}. The unitary matrix $U_j$ is constructed by columnwise arranging the left ($j=L$) or the right ($j=R$) eigenvectors of the states with $\Re(E)<0$, and $n_{\rm al}=(1/2) \; \sum_{\bf r} x y /\ell^2$ represents $n$ in the atomic limit and at half-filling. Here $\ell$ is the linear dimension of the system in $x$ and $y$ directions. For all four possible combinations of (a) $j$ and $k$, and (b) ${\boldsymbol h}$, we obtain identical phase diagrams of $H_{\rm SO}$ [Fig.~\ref{fig:Fig4}(c)]. For small and moderate NH couplings, $Q^{\rm NH}_{xy}$ takes quantized values $0.5$ and $0.0$, respectively representing NH quadrupole and trivial insulators. The former features intrinsic second-order ISE. By contrast, $Q^{\rm NH}_{xy}$ is not well defined for large $h$, similar to the situation with $B_{\rm NH}$.

{\bf Majorana ISE}.~NH SC fractal hosting topological superconductivity, can harbor ISEs for neutral Majorana fermions. To demonstrate first-order Majorana ISE, we account for the only local or onsite pairing available to the spinless system. The total Hamiltonian then reads 
\begin{equation}~\label{eq:Majoranafirstorder}
H^{\rm pair}_{\rm FO}=H_{\rm FO} + \Delta_{\rm FO} \sum_{j} \left[ c^\dagger_{j} \tau_2 c^\dagger_{j} + c_{j} \tau_2 c_{j} \right],
\end{equation} 
where $\Delta_{\rm FO}$ is the (real) amplitude of the first-order topological pairing. This model possesses an exact particle-hole symmetry as shown in Supplementary Note 3. In the absence of NH coupling (${\boldsymbol h}=0$) the above pairing can sustain one-dimensional edge modes of neutral Majorana fermions. With the addition of the NH coupling $h_x$, for example, the paired state shows skin effect at outer left edge of the SC with OBC. And with PBC only in the $x$ direction, the skin effect appears at the inner left edges of the fractal lattice, yielding first-order Majorana ISE [Fig.~\ref{fig:Fig5}(a)]. We also construct a global phase diagram of the above NH paired state model in the $(m_0, h_x)$ plane by generalizing the NH Bott index for Nambu doubled fermionic degrees of freedom [Fig.~\ref{fig:Fig5}(b)]. It features three NH paired states with $B_{\rm NH}=0$ (trivial), $-1$ and $-2$ (both being topological). When $B_{\rm NH}$ acquires nontrivial integer values, Majorana ISEs are intrinsic in nature.

By the same token, second-order Majorana ISE can be showcased by introducing a local higher-order topological pairing of (real) amplitude $\Delta_{\rm SO}$ among spinful fermions on the NH SC fractal, with the total Hamiltonian 
\begin{equation}~\label{eq:Majoranasecondorder}
H^{\rm pair}_{\rm SO}=H_{\rm SO} + \Delta_{\rm SO} \sum_{j} \left[ c^\dagger_{j} \sigma_3 \tau_2 c^\dagger_{j} + c_{j} \sigma_3 \tau_2 c_{j} \right],
\end{equation} 
possessing exact particle-hole symmetry~\cite{mannanandyroy2021:frac, broysoloHOTSC2020, royjuricicoctupole2021}, as shown in Supplementary Note 4. When ${\boldsymbol h}=0$, $H^{\rm pair}_{\rm SO}$ supports four corner localized zero-energy Majorana modes. For ${\boldsymbol h}=(\pm h, \pm h, 0)$ with $h>0$, the paired state displays outer corner skin effect on the SC fractal with OBC. On the other hand, when PBCs are imposed in the $x$ and $y$ directions, the system shows second-order Majorana ISE around the inner corners of the fractal lattice [Fig.~\ref{fig:Fig5}(c)]. A global phase diagram of $H^{\rm pair}_{\rm SO}$ can be constructed by computing the NH quadrupole moment for Nambu doubled fermions [Fig.~\ref{fig:Fig5}(d)]. It supports NH paired states with $Q^{\rm NH}_{xy}=0.0$ (trivial) and $0.5$ (topological). The latter features intrinsic second-order Majorana ISE.

\noindent 
{\bf Discussions} \\
To summarize, here we show that the inner edges and corners of the SC fractal lattices with PBC, respectively depict first- and second-order skin effects of both charged and neutral Majorana fermions: A phenomenon, which we dub as ISE. These outcomes are insensitive to the generation number of the SC fractal lattice, as shown in the Supplementary Figs.~7 and 8. With increasing generation number first-order (second-order) ISE slightly spreads away from the edges (corners) of the center hollow part of the fractal lattice, and starts to appear on newly emerging inner boundaries. The hierarchy of fractal generation this way manifests on the NH ISEs. Even though in the main manuscript we show evidence of ISEs inside topological phases with $B_{\rm NH}=-1$ and $Q^{\rm NH}_{xy}=0.5$, for example, the ISEs are insensitive to the NH topological invariant and can be seen anywhere in the phase diagrams. See Supplementary Figs.~9 and 10. While the intrinsic first-order and second-order ISEs are characterized by different quantized topological invariants, namely $B_{\rm NH}$ and $Q^{\rm NH}_{xy}$, they can only be distinguished geometrically (appearing around inner edges and corners, respectively) when these invariants are trivial and not well defined (as in the case with point gap). Possible realization of two-dimensional NH Chern insulators with $B_{\rm NH}=-1$ and the associated ISE from higher-dimensional topological phases via holographic duality stands as an interesting avenue to pursue in the future, since such mapping exists in Hermitian systems~\cite{choryuqi:holoraphy2016}.

Existence of the ISE in the thermodynamic limit can be established quantitatively from the scaling of the fraction of all the wavefunctions (left or right) localized at the innermost edge (for the first-order ISE), denoted by $f^{\rm FO}_{\rm ISE}$, with the generation number or number of sites ($N$) of the fractal lattice. We find that $f^{\rm FO}_{\rm ISE}$ always saturates to a finite value as $N \to \infty$. In addition, we also find that $f^{\rm FO}_{\rm ISE}$ in a given generation of the Sierpinski carpet fractal lattice always remains finite for any value of the mass parameter ($m_0$) and the NH coupling ($h_x$), see Eq.~(\ref{eq:HamilNHfirstorder}), encompassing all three phases appearing in the phase diagram Fig.~\ref{fig:Fig3}(d): NH Chern (trivial) insulator with $B_{\rm NH}=-1 \; (0)$ and NH insulator where $B_{\rm NH}$ is not well defined. Only near the phase boundary between two topologically distinct insulating phases it shows a \emph{weak} deep as then the system becomes gapless. Similar conclusions also hold for the second-order ISE. These findings are displayed in Supplementary Fig.~11.

The proposed ISEs on periodic fractal lattices should be contrasted with dislocation NH skin effect with PBC~\cite{panigrahi:2021, schindlerprem:2021, Moghaddam:PRB2021}. Dislocation NH skin effect crucially depends on the relative orientation between the associated Burgers vector and the direction of the skin effect at outer boundaries with open geometry. A skin effect appears at the defect core under PBC only when they are orthogonal to each other~\cite{panigrahi:2021}. Furthermore, such dislocation skin effects can be masked by localized dislocation bound states, making it challenging to detect in experiments. By contrast, the ISE on periodic fractal lattices is a generic phenomenon, which can be either first-order or second-order in nature, and should be detected in experiments.

We notice when inner boundaries are externally engineered on a square lattice, amorphous network and Ammann-Beenker quasicrystals, conforming to Corbino geometry, they can at least display first-order ISE when periodic boundary condition is imposed along only the horizontal $x$ direction. See Supplementary Fig.~12. This observation should be contrasted with the ISE at the inner boundary of fractal lattices, which is an intrinsic geometric property (not generated artificially from the outset), stemming from its self-similarity symmetry. Additional numerical results shown in various Supplementary figures are discussed in Supplementary Note 5.

For Majorana skin effect we consider the amplitudes of topological pairings to be real, which can in principle be accommodated by either electron-phonon interaction or proximity effects on parent NH quantum fractal lattices (constituting the normal state) at low temperatures. This way neutral Majorana fermions inherit non-Hermiticity from the normal state and produce skin effect in the paired state. Alternatively, the parent state can be considered to be Hermitian, while the pairing amplitudes are non-Hermitian. Although microscopic origin of such situation is presently unknown, it nonetheless can also produce NH Majorana skin effects~\cite{tanmoy:review}. A complete self-consistent analysis of the pairing amplitude in the presence of system-to-environment interaction and effective attraction among electrons (either proximity or phonon mediated) remains an open challenging problem for future investigations. Among electronic systems, designer quantum~\cite{cmsmith2019:frac} and molecular~\cite{wu2015:frac} materials are the most prominent ones where desired fractal lattices and requisite hopping elements can be engineered, and our predictions on ISEs for charged and Majorana fermions can be tested, when they interacts with environment, yielding gain and/or loss, captured by non-Hermitian operator(s).

Even though a one-to-one correspondence between the nature of the system-to-environment interaction and the resulting NH operator remains elusive, classical metamaterials, such as photonic and phononic lattices, and topolectric circuits, on the other hand, constitute a promising platform, where desired NH operators can be engineered directly and the proposed ISEs can be measured experimentally. On photonic lattices with gain and/or loss, yielding a NH setup, topological edge modes, bulk exceptional points and bulk Fermi arcs connecting them have already been reported~\cite{NH-photonic:1, NH-photonic:2, NH-photonic:3, NH-photonic:4, NH-photonic:5, NH-photonic:6}. In this setup the ISE can be detected via two point pump probe~\cite{defect-photonic:1} or reflection spectroscopy~\cite{defect-photonic:2} around the inner boundaries of photonic fractal lattices with PBC. NH topology has also been revealed on mechanical systems~\cite{NH-mechanical:1, NH-mechanical:2, NH-mechanical:3}, where ISE can be unveiled by directly measuring the right eigenvectors, for example~\cite{NH-mechanical:3}. On topolectric circuits, requisite NH couplings can be tailored by suitable resistance arrangements and the desired ISEs can be measured from complex impedance~\cite{NH-topolectric:1, NH-topolectric:2, NH-topolectric:3, NH-topolectric:4}. In this setup a Hermitian quasicrystalline quadrupolar insulator has been realized~\cite{topolectric:quadrupole}. Thus ISEs can be demonstrated on periodic NH fractal topolectric circuits.

Our numerical simulations suggest that to observe the ISE on SC fractals (a) hopping elements only up to the next-nearest-neighbor sites along the principal axes directions (with $R=2 a$) are sufficient and (b) PBCs are required in one (for first-order ISE) or two (for second-order ISE) directions. Therefore, experimental observations of ISEs will require only a limited number of additional couplings, especially among the waveguides (on photonic lattices) or mechanical resonators (on phononic lattices) or electrical nodes (on topolectric circuits), playing the role of lattice sites, otherwise residing near the outer edges of the underlying SC fractal metamaterials. Moreover, recently longer range hopping has been engineered on topolectric circuits~\cite{topolectric:longrange}, and SC fractal lattices have been realized on metamaterials, harboring Hermitian first- and second-order topological phases~\cite{sierHOTExp1, sierHOTExp2, sierHOTExp3}. These realistic requirements and recent achievements make our proposal promising, and most vitally within the reach of the existing experimental facilities.

The present expedition also constitutes foundations of several fascinating avenues through the landscape of NH topology and skin effect, among which ISEs and its dependence on the internal geometry on other fractal lattices, such as Sierpinski triangle representing a homogeneous fractal, is a promising one. Furthermore, Fig.~\ref{fig:Fig2} displaying an interplay between the eight-fold rotational symmetry of the Ammann-Beenker tiling and its overall octagonal shape for NH skin effect, serves as the torchbearer leading to an unexplored territory of NH topological phases on quasicrystals. These fascinating directions will be explored systematically in future.

\noindent 
{\bf Methods}\\
{\bf Hamiltonian construction}.~To demonstrate the inner skin effect (ISE) on Sierpinski carpet fractal lattices with periodic boundary condition and their connections with non-Hermitian (NH) topology we focus on a number of toy models, namely $H_{\rm FO}$ [Eq.~(\ref{eq:HamilNHfirstorder})] and $H_{\rm SO}$ [Eq.~(\ref{eq:HamilNHsecondorder})], respectively capturing first-order and second-order NH skin effects for charged fermions, as well as $H^{\rm pair}_{\rm FO}$ [Eq.~(\ref{eq:Majoranafirstorder})] and $H^{\rm pair}_{\rm SO}$ [Eq.~(\ref{eq:Majoranasecondorder})] capturing first-order and second-order NH skin effects for neutral Majorana fermions, respectively. Each real space Hamiltonian is constructed from its momentum space counterpart in the following way. We analyze transformations of each term appearing in the Bloch Hamiltonian under the pertinent discrete symmetry transformations, such as reflections and rotations, and identify the corresponding term in the real space that transforms identically under all the symmetry operations. Subsequently, all the terms in the Bloch Hamiltonian gets replaced by their real space cousins. This analysis is shown in details in the Supplemental Information. Finally we augment all the short-range hopping elements by exponentially decaying but finite range hopping in terms of a function shown in Eq.~(\ref{eq:hoppingconfiguration}) such that all the sites on the fractal lattice are well connected. Identical prescription is implemented on a square lattice, Ammann-Beenker quasicrystal and amorphous network.

{\bf Inner skin effect}.~Once the stage is set, we numerically diagonalize each Hamiltonian for a wide range of parameters. To capture the conventional NH skin effect we impose open boundary condition in all directions in all the systems. The first-order and second-order conventional NH skin effects manifest accumulation of a large number of eigenvectors (left or right) of the corresponding Hamiltonian respectively near a specific edge and corner of the system. In order to showcase the NH ISE we impose periodic boundary condition in one and both directions for these two cases, respectively. For the first-order ISE the periodic boundary condition is imposed in the direction along which the systems displays conventional NH skin effect in open boundary system. Note that both first-order and second-order ISE is exclusively available on Sierpinski carpet lattice with requisite periodic boundary condition (mentioned above) for which a macroscopic number of eigenvectors (left or right) accumulate near one of the innermost edge and corner of the system, respectively.

{\bf Global phase diagram}.~To construct the global phase diagram of these models, we consider the half-filled ground state, constructed by filling all the state with $\Re(E)<0$, where $E$ is the complex energy eigenvalue. Respectively the first-order and second-order ISE can be tied with the quantized NH Bott index ($B_{\rm NH}$) and NH quadrupole moment ($Q^{\rm NH}_{xy}$), which we compute on such a ground state. Additional details related to the computation of these invariants are discussed in the manuscript.

\noindent 
{\bf Data Availability}\\
The data for generating the figures presented in the main text and Supplementary Information are available upon reasonable request from the corresponding author and Sourav Manna (sourav4phy@gmail.com).

\noindent 
{\bf Code Availability}\\
Software code in Pyhton for generating all the results presented in the main text and Supplementary Information are available from the corresponding author and Sourav Manna (sourav4phy@gmail.com) upon reasonable request.

\noindent 
{\bf Acknowledgments}\\
S.M. thanks Weizmann Institute of Science, Israel Deans fellowship through Feinberg Graduate School for financial support. B.R. was supported by the Startup grant from Lehigh University. 

\noindent
{\bf Author contributions}\\ 
S.M. performed all the numerical calculations. B.R. conceived and structured the project, and wrote the manuscript.

\noindent 
{\bf Corresponding Author}\\
Bitan Roy (bitan.roy@lehigh.edu).

\noindent 
{\bf Conflict of interests}\\
The authors declare no conflicts of interest.


\end{document}